\documentclass[12pt]{article}
\usepackage{bm}
\usepackage{amsmath}
\usepackage{amssymb}
\usepackage{graphicx}
\usepackage{amsmath}
\usepackage{bbm}

\begin{document}

\baselineskip 15pt

\title{\bf Entanglement, Nonlocality, Superluminal Signaling and Cloning}

\author{GianCarlo Ghirardi\footnote{e-mail: ghirardi@ts.infn.it}\\ {\small
Emeritus, University of Trieste,}\\ {\small The Abdus
Salam International Centre for Theoretical Physics, Trieste.}  }

\date{}

\maketitle

\section{Introduction}

Entanglement has been considered by E. Schr\"{o}dinger  [1] as: {\it The most characteristic trait of Quantum Mechanics, the one which enforces its entire departure from classical lines of thought}. Actually, the just mentioned unavoidable departure from the classical worldview raises some serious problems when entanglement of far away quantum systems is considered in conjunction with the measurement process on one of the constituents. These worries have been, once more, expressed with great lucidity by Schr\"{o}dinger himself [1]: {\it It is rather discomforting that the theory should allow a system to be steered or piloted into one or the other type of state at the experimenter`s mercy, in spite of his having no access to it}.

	All those who are familiar with quantum theory will have perfectly clear the formal and physical aspects to which the above sentences make clear reference: they consist in the fact that, when dealing with a composite quantum system whose constituents are entangled and far apart, the free will choice of an observer to perform a measurement at one wing of the apparatus and the quantum reduction postulate imply that the far away state ``jumps`` in a state which depends crucially from the free will choice of the observer performing the measurement and from the random outcome he has got. Just to present a quite elementary case, let us consider a quantum composite system $S=S^{(1)}+S^{(2)}$, in an entangled state $\vert\psi(1,2)\rangle$:
\begin{equation}
\vert\psi(1,2)\rangle=\sum_{i}p_{i}\vert\phi_{i}^{(1)}\rangle\otimes\vert\gamma_{i}^{(2)}\rangle,\;\;p_{i}\geq 0,\;\;\sum_{i}p_{i}=1.
\end{equation}
In this equation (the Schmidt biorthonormal decomposition) the sets   $\{\vert\phi_{i}^{(1)}\rangle\}$ and  $\{\vert\gamma_{i}^{(2)}\rangle\}$ are two orthonormal sets of the Hilbert spaces of system $S^{(1)}$ and $S^{(2)}$, respectively, and, as such, they are eigenstates of appropriate observables $\Phi^{(1)}$ and  $\Gamma^{(2)}$ of such subsystems. Suppose now that subsystem $S^{(2)}$ is subjected to a measurement of the observable    $\Gamma^{(2)}$  and suppose that in the measurement the outcome   (one of its eigenvalues) $\Gamma^{(2)}=g_{r}$ is obtained. Then, reduction of the wave packet leads instantly to the state  $\vert\phi_{r}^{(1)}\rangle\otimes\vert\gamma_{r}^{(2)}\rangle$ for which one can claim that if system $S^{(1)}$ is subjected to a measurement of the observable $\Phi^{(1)}$ the outcome  $\Phi^{(1)}=f_{r}$ will occur with certainty. Since this outcome, before the measurement on system $S^{(2)}$ had a nonepistemic probability  $p_{r}^{2}$ of occurrence, one can state that the observation of $\Gamma^{(2)}$   has caused the instantaneous emergence at-a-distance of a definite property (which, according to quantum mechanics, one cannot consider as possessed in advance) of subsystem $S^{(1)}$, i.e. the one associated to the eigenvalue  $f_{r}$ of the observable $\Phi^{(1)}$. An analogous argument can  obviously be developed without making reference to the Schmidt decomposition but to an arbitrary measurement on subsystem $S^{(2)}$, and will lead, in general, to the emergence of a different property of subsystem $S^{(1)}$ (typically the outcome of another appropriate measurement on this system becomes certain).

	The just described process makes perfectly clear the nonlocal character of quantum mechanics, a fact that subsequently  has been  precisely identified by the illuminating work of J.S. Bell [2].
	
	The situation we have just described will allow the reader to understand how it has given rise to the so called problem of faster-than-light signaling. If my action on system $B$, which takes place and is completed at a space like separation from system $A$, affects this system making instantaneously actual one of its potentialities, I can hope to be able to take advantage of this quantum peculiarity to make the observer  $A$ aware of the fact that I am performing some precise action on subsystem $B$ at a space-like separation from him.
	
	And, actually, this is what happened. From the seventies up to now, an innumerable set of proposals of taking advantage of entanglement and the reduction of the wave packet to achieve superluminal communication between distant observers appeared in the literature, proposals aiming to exploit this exciting possibility and to put into evidence  the incompatibility  of quantum mechanics with special relativity. Fortunately, as I will show in this paper, all proposal advanced so far, and, in view of some general theorems I will discuss below, all conceivable proposals of this kind, can been proven to be basically flawed in a way or another.
	
	This chapter is devoted to discuss this important and historically crucial aspect of modern physics. As such, it has more an historical than a research interest. However, I believe that the reconsideration of the debate about this issue will be useful for the reader since many not so well known and subtle aspects of quantum mechanics will enter into play. 
	
	Before coming to a sketchy outline of the organization of the whole paper I would like to call the attention of the reader to a quite peculiar fact. When the so-called quantum measurement problem arose and was formalized by J. von Neumann [3], the attention of the scientific community was not concentrated on the possible conflicts between quantum mechanics and relativity; quantum mechanics was considered as a fundamentally nonrelativistic description of natural processes. Obviously, everybody had clear that the problem of its relativistic generalization had to be faced, but the debate concerned the nonrelativistic aspects of the theory and nobody had raised the question of possible conflicts between the two pillars of modern science\footnote{For this one should wait  the celebrated EPR paper, which appeared, just as  Sch\"{o}dinger's paper [1] in which the instantaneity of the reduction is seen as problematic, 2 years after von Neumann's precise formalization of the effect of a measurement.}. In spite of that, the cleverly devised prescription of wave packet reduction, which was elaborated without having in mind relativistic potential oddities, turned out to be such that, in spite of its nonlocal nature and instantaneity, it did not allow to violate the basic relativistic request of no-faster-than-light signaling.
	
	A brief outline of the organization of the chapter follows. After recalling the relevant aspects of the way in which quantum mechanics accounts for natural processes, we will describe various  proposals for achieving faster-than-light signaling which have been put forward, and point out the reasons for which they are basically incorrect. To conclude this part we will present a general theorem showing that quantum mechanics, in its standard version, cannot in principle lead to superluminal communication.
	
	However, the most interesting part of the debate is not the one we have just mentioned. In 1982 an analogous but quite different proposal of faster-than-light signaling has been put forward by N. Herbert [4]. The idea consisted in taking, as usual, advantage of the entanglement of far away subsystems and of wave packet reduction, but a new device was called into play: an hypothetical machine which could perform the task of creating many copies of an arbitrary state of a quantum system (a sort of ``quantum xeroxing  machine"). The interesting point it that at that time no general argument had been developed proving this task impossible. So, the mistaken suggestion of Herbert triggered the derivation of a theorem, the so-called no-cloning theorem, which was not known and which represents a quite relevant achievement which stays at the very basis of many important recent developments  and which, besides proving that Herbert's proposal was  unviable, plays a fundamental role for quantum cryptography and  quantum computation.
	
	As it is obvious, an hypothetical quantum device allowing faster-than-light communication would give rise to a direct and serious conflict with the special theory of relativity. As already stated, such a device is excluded by quantum mechanics. This, however, does not eliminate completely the potential tension of the nonlocal nature of quantum theory with the basic principles of relativity theory. The central issue is that the instantaneous collapse of the statevector of the far away system, even if it cannot be used to transfer energy or information at a superluminal speed, indicates that, in a way or another, an action performed in a given space-time region has some ``effect" on systems at a space-like separation. Einstein has qualified this aspect of the theory as  ``a spooky action at-a-distance" which he could not accept. A. Shimony, by stressing the fact that the theory cannot be used to actually communicate superluminally has expressed his opinion that there is some sort of  ``peaceful coexistence of quantum mechanics and relativity" and has suggested to speak, in place of an ``action" of a ``passion" at-a-distance, to stress the peculiar nature of the perfect correlations of the outcomes, which, before any measurement process, individually have a fundamentally random nature, i.e., only a certain probability of occurrence. 
	
	Recently, the just mentioned problem has seen a revival due to the elaboration of the so-called ``collapse models", i.e.,  modifications of quantum mechanics which, on the basis of a unique, universal dynamical principle account both for the quantum evolution of microscopic systems as well as for the reduction process when macroscopic systems enter into play. Such theories, the best known of which is the one presented in ref.[ 5] usually quoted as ``The GRW Theory", have been worked out with the aim of solving the macro-objectification or measurement problem at the nonrelativistic level, and the fact that they get the desired result in a clean, mathematically rigorous and conceptually precise way has raised the interest of various scientists, among them the one of Bell [6-8]. After the complete formalization of such approaches, it has been natural to start investigating whether they admit relativistic generalizations. Since they, agreeing with the quantum predictions concerning microsystems, exhibit (essentially) the same nonlocal aspects as standard quantum mechanics, the question of wether they actually can be made compatible with relativity has attracted a lot of attention. The serious work of various physicists in recent years has made clear that the program can be pursued, which means that one can have a theory inducing instantaneous collapses at-a-distance which does not violate any relativistic request. We consider it interesting to devote the conclusive part of this chapter to outline the investigations along these lines   and to discuss their compatibility with the principles of special relativity. 
	
	\section{The relevant formal aspects of the theory}
	
	\subsection{The general rules}
	
	As is well known, quantum mechanics asserts that the most accurate specification of the state of a physical system is given by the statevector $\vert\Psi\rangle$, an element of the Hilbert space $\cal{H}$ associated to the system itself. When one deals with a statistical ensemble of identical systems, an equivalent and practical mathematical object is the statistical operator $\rho$ which is the weighted sum of the statistical operators $\vert\psi_{i}\rangle\langle\psi_{i}\vert$ corresponding to the pure states $|\psi_{i}\rangle$ of the members of the ensemble: $\rho=\sum_{i}p_{i} \vert\psi_{i}\rangle\langle\psi_{i}\vert$, with $p_{i}\geq0,\;\sum_{i}p_{i}=1$. For an homogeneous ensemble or an individual system in a pure state $\vert\Psi\rangle$, the statistical operator is a projection operator: $\rho=\vert\Psi\rangle\langle\Psi\vert$. 
	
	The observables of the theory are represented by self-adjoint operators of the Hilbert space $\cal{H}$, which are characterized by their eigenvalues and eigenvectors. For the observable $\Omega$ one writes its eigenvalue equation as \footnote{For simplicity we will deal with observables with a purely discrete spectrum, the changes for the continuous case been obvious.}:   
\begin{equation}
\Omega\vert\omega_{k,\alpha}\rangle=\omega_{k}\vert\omega_{k,\alpha}\rangle,
\end{equation}
where the index $\alpha$ is associated to the possible degeneracy of the eigenvalue $\omega_{k}$. A crucial feature implied by the assumption of self-adjointness of the operators representing physical observables is that their spectral family, i.e. the projection operators $P_{r}=\sum_{\alpha}\vert\omega_{r,\alpha}\rangle\langle\omega_{r,\alpha}\vert$ on their eigenmanifolds, correspond to a resolution of the identity: $\sum_{r}P_{r}=I$, $I$ being  the identity operator on $\cal{H}$.

For what concerns the physical predictions, it is stipulated that in the case of a system in a pure state $\vert\Psi\rangle$ one has to express it as a linear combination of the eigensates of the observable (let us call it $\Omega$) corresponding to the microscopic physical quantity which one intends to measure: $\vert\Psi\rangle=\sum_{k,\alpha}c_{k,\alpha}\vert\omega_{k,\alpha}\rangle$. Then the theory asserts that the probability $P(\Omega=\omega_{r}\vert\Psi)$ of getting the outcome $\omega_{r}$ in the measurement of $\Omega$ when the system is in the pure state $\vert\Psi\rangle$, is given by $\sum_{\alpha}\vert c_{r,\alpha}\vert^{2}$, a quantity which coincides with the square of the norm $\vert\vert P_{r}\vert\Psi\rangle\vert\vert^{2}$ of the projection of the state onto the relevant eigenmanifold . This rule becomes, in the statistical operator language, $P(\Omega=\omega_{r}\vert\rho)=Tr[P_{r}\rho]$, where the symbol $Tr$ means that  the sum of the diagonal elements of the quantity in square brackets in an arbitrary orthonormal complete basis of $\cal{H}$ must be taken (this sum is easily proved not to depend on the chosen basis). Note that, using the complete set of the eigenstates of an operator $\Omega$ to evaluate the Trace, one immediately sees that  its quantum average  can be simply expressed as $\langle\Omega\rangle=Tr[\Omega\rho]$. It is an important mathematical fact that the Trace operation is linear and enjoys of the following formal feature: given two arbitrary (bounded) operators $\Lambda$ and $\Gamma$ of $\cal{H}$, $Tr[\Lambda\Gamma]=Tr[\Gamma\Lambda]$. 

Before concluding this subsection we must also mention the effect on the statevector of performing a measurement process. Actually, two kinds of measurements can be carried out: the nonselective and the selective ones, i.e. those in which one measures an observable without isolating the cases in which a precise eigenvalue is obtained or, alternatively, those in which one is interested only in a definite outcome. They are represented, in the statistical operator language, by the two following formal expressions:
\begin{equation}
\rho_{before}\rightarrow\rho_{after}\equiv \sum_{k}P_{k}\rho_{before}P_{k},
\end{equation}
\begin{equation}
\rho_{before}\rightarrow\rho_{after}\equiv P_{k}\rho_{before}P_{k}/Tr[P_{k}\rho_{before}],
\end{equation}

with obvious meaning of the symbols.

\subsection{Composite systems}
From now on we will be mainly interested in dealing with  quantum systems $S$ composed of two constituents, $S^{(1)}$ and $S^{(2)}$. Accordingly, their statevector $\vert\Psi(1,2)\rangle$ is an element of the tensor product ${\cal H}^{(1)}\otimes {\cal H}^{(2)}$ of the Hibert spaces of the constituents. As is well known, in the considered case two radically different situations may occur: in the first one the statevector is simply the direct product of  precise statevectors for the constituents $\vert\Psi(1,2)\rangle=\vert\phi(1)\rangle\otimes\vert\gamma(2)\rangle$, and in such a case both constituents possess precise physical properties; alternatively, the statevector is entangled, i.e., it cannot be written in this form but it involves the superposition of factorized states, typically  $\vert\Psi(1,2)\rangle=\sum_{i}c_{i}\vert\phi_{i}(1)\rangle\otimes\vert\gamma_{i}(2)\rangle$.

An extremely important point concerning composite systems is the following. Suppose one has a composite system and he is interested only in the outcomes of perspective measurement processes on  one of the constituents. Then, one can easily convince himself that the simplest way of dealing with this problem is to consider the reduced statistical operator $\tilde{\rho}(1)$, obtained by taking the partial trace of the full statistical operator on the Hilbert space of the subsystem $S^{(2)}$ one is not interested in. At this point, to evaluate the probability of the outcomes of measurements of observables of the system of interest $S^{(1)}$, one  can use the reduced statistical operator and the same prescriptions we have used for the general case \footnote{An elementary way to see this is to evaluate the probabilities of the joint outcomes of the measurement of a pair of observables, one for system $S^{(1)}$ and one for  $S^{(2)}$, and then to sum  on all possible outcomes of the measurement on the system we are not interested in.} :
\begin{eqnarray}
\tilde{\rho}(1)&=&Tr^{(2)}[\rho(1,2)];\;\;P(\Phi^{(1)}=f_{r}\vert\rho(1,2))=Tr^{(1+2)}[P_{r}^{(1)}\rho (1,2)]\equiv Tr^{(1)}[P_{r}^{(1)}\tilde{\rho}(1)]; \nonumber \\
 \langle\Phi^{(1)}\rangle&=& Tr[\Phi^{(1)}\tilde{\rho}(1)].
\end{eqnarray}
It goes without saying that the operator $P_{r}^{(1)}$ in the previous equation is the projection operator onto the linear eigenmanifold associated to the eigenvalue $f_{r}$ of $\Phi^{(1)}$.

\subsection{von Neumann's  ideal measurement scheme  and its limitations}

For the subsequent analysis it is important to briefly recall the so-called Ideal Measurement Scheme introduced by von Neumann in his celebrated book, ref.[3], and its limitations. The idea is quite simple: we are interested in ``measuring" a microscopic observable, which is not directly accessible to our senses. Suppose then we have  a microsystem $s$ in a state $\vert\varphi_{i}^{(s)}\rangle$ which is in an eigenstate of a micro-observable $\Sigma^{(s)}$ pertaining to the eigenvalue $s_{i}$. How can one ascertain such a value, which, if a  measurement is performed, according to the quantum rules will be obtained with certainity ? Von Neumann assumed that there exist a macroscopic object $M$ which can be prepared in a ready state $\vert m_{0}\rangle$ and can be put into interaction with the microsystem. The interaction leaves unaltered the microstate while it induces, in a quite short time interval, the following evolution of the microsystem+apparatus:
\begin{equation}
\vert \varphi_{i}\rangle\otimes\vert m_{0}\rangle\rightarrow \vert \varphi_{i}\rangle\otimes\vert m_{i}\rangle,
\end{equation}
where the states $\vert m_{i}\rangle$ are assumed to be orthogonal ($\langle m_{i}\vert m_{j}\rangle=\delta_{i,j}$), macroscopically and perceptively different (typically  they are associated to different locations of the pointer of the macro-apparatus). Then, an observer, by looking at the measuring apparatus gets immediately the desired information concerning the value ($s_{i}$) of the microvariable. 

The scheme is usually qualified as ideal because, in practice, the final apparatus states are not perfectly orthogonal and because very often the state of the microsystem  is disturbed (or even the system is absorbed) in the measurement. The just mentioned scheme has an immediate important implication; the validity of Eq.(6) and the linear nature of  Schr\"{o}dinger's evolution equation imply that if one triggers the macroapparatus in its ready state with a superposition of the eigenstates of $\Sigma^{(s)}$, one has:
\begin{equation}
\sum_{i}c_{i}\vert \varphi_{i}\rangle\otimes\vert m_{0}\rangle\rightarrow \sum_{i} c_{i}\vert \varphi_{i}\rangle\otimes\vert m_{i}\rangle,
\end{equation}
which is an entangled state of the microsystem and the macroapparatus.

Eq.(7)  has given rise to one of the most debated problems of quantum mechanics, the so-called measurement or macro-objectification problem. In fact its r.h.s. corresponds to an entangled state of the system and the apparatus and in no way whatsoever to a state corresponding to a precise outcome\footnote{Note that in ref.[9] it has been proven that the occurrence of the embarrassing superpositions of macroscopically different states does not require that the measurement proceeds according to the ideal scheme of von Neumann. The same conclusion can be derived as a consequence of the necessary request that quantum mechanics governes the whole process and that one can perform a reasonably reliable measurement  ascertaining the microproperty of the measured system.} . The orthodox way out from this puzzle consists in resorting to the postulate of wave packet reduction: when a superposition of different macrostates emerges, a sudden change of the statevector occurs, so that one has to replace the r.h.s. of the previous equation with one of its terms, let us say $\vert \varphi_{j}\rangle\otimes\vert m_{j}\rangle$. This specific reduction occurs with  probability $\vert c_{j}\vert^{2}$. We will not enter, here, in this deep debate, we  simply mention that it amounts to accept (as many scientists did) that the linear character of the theory is violated (the reduction process is nonlinear and stochastic while the quantum evolution is linear and deterministic) at an appropriate ( but not precisely specified) macroscopic level. Incidentally, von Neumann himself has proposed that the transition from the superposition to one of its terms occurs when a conscious observer becomes aware of the outcome (reduction by consciousness). Recently, various proposals of theories which, on the basis of a unique dynamical principle, account both for the linear nature of the evolution at the microscopic level as well as for the discontinuous changes (collapses) occurring when macrosystems are involved, have been put forward. We refer the reader to ref.[10] for an exhaustive analysis of such model theories.

\subsubsection{Limits to the ideal scheme due to additive conservation laws}

There are   limitations to the von Neumann ideal scheme that we must mention because they have played a role in the refutation of some proposals of faster-than-light communication. Such  limitations have been identified  by Wigner [11], Araki and Yanase [12,13] in a series of interesting papers and subsequently they have been generalized in refs.[14,15]. The analysis by these authors takes into account the existence of additive conserved quantities for the system+apparatus system to derive precise conditions on a process like the one of Eq.(6) which stays at the basis of the von Neumann treatment. Let us summarize the procedure in a sketchy way. The process described by Eq.(6) represents the unitary evolution of the system+apparatus during the measurement process. Let us therefore write it as: $U\vert\varphi_{i}, m_{0}\rangle=\vert\varphi_{i},m_{i}\rangle$.  Let us suppose that there exists an additive conserved quantity $\Gamma=\gamma^{(s)}\otimes I^{(A)}+I^{(s)}\otimes\gamma^{(A)}$ of the whole system and let us evaluate the matrix element of $\Gamma$, $\langle\varphi_{i},m_{0}\vert\Gamma\vert\varphi_{j},m_{0}\rangle$ by taking into account that $\Gamma$ commutes with $U$, which implies :
\begin{equation}
\langle\varphi_{i},m_{0}\vert(\gamma^{(s)}\otimes I^{(A)}+I^{(s)}\otimes\gamma^{(A)})\vert\varphi_{j}m_{0}\rangle=\langle\varphi_{i},m_{0}\vert U^{\dag}(\gamma^{(s)}\otimes I^{(A)}+I^{(s)}\otimes\gamma^{(A)})U \vert\varphi_{j}m_{0}\rangle.
\end{equation}

 We then have:
\begin{eqnarray}
& &\langle\varphi_{i},m_{0}\vert(\gamma^{(s)}\otimes I^{(A)}+I^{(s)}\otimes\gamma^{(A)})\vert\varphi_{j}m_{0}\rangle=\langle\varphi_{i}\vert\gamma^{(s)}\vert\varphi_{j}\rangle+\delta_{ij}\langle m_{0}\vert\gamma^{(A)}\vert m_{0}\rangle= \nonumber \\
& &\langle\varphi_{i},m_{0}\vert U^{\dag}(\gamma^{(s)}\otimes I^{(A)}+I^{(s)}\otimes\gamma^{(A)})U \vert\varphi_{j}m_{0}\rangle = \nonumber \\
& &\langle\varphi_{i},m_{i}\vert(\gamma^{(s)}\otimes I^{(A)}+I^{(s)}\otimes\gamma^{(A)})\vert\varphi_{j},m_{j}\rangle=\nonumber \\
& &\delta_{ij}\langle\varphi_{i}\vert\gamma^{(s)}\vert\varphi_{j}\rangle+\delta_{ij}\langle m_{i}\vert\gamma^{(A)}\vert m_{j}\rangle.
\end{eqnarray}

Comparison of the final expression with the one after the equality sign in the first line shows that, in the considered case, one must have, for $i\neq j$, $\langle\varphi_{i}\vert\gamma^{(s)}\vert\varphi_{j}\rangle=0$ which amounts to the condition that the observable $\Sigma^{(s)}$ which we want to measure on the microsystem must commute with the microsystem part $\gamma^{(s)}$ of the conserved additive quantity. If this is not the case (as it happens when $\Sigma^{(s)}$ is a component of the angular momentum of the system which does not commute with the other components), a process like the one of Eq.(6) turns out to be impossible; terms must be added to the r.h.s. involving other states of the microsystem besides $\vert\varphi_{i}\rangle$ and also other states of the apparatus. In refs.[13-15] it has been shown that in order to go as near as possible to the ideal case one must make more and more large the square of the norm of the state $\gamma^{(A)}\vert m_{0}\rangle$. In the case of an angular momentum measurement this means to make  the mean value of the square of the angular momentum component extremely large. Actually, in the case of the measurement of the spin component of a spin 1/2 particle, the ``distorsion" of the state by the measurement, a quantity which can be estimated by the squared norm $\epsilon^{2}$ of the state which has to be added to the r.h.s. of Eq.(6), must satisfy: $\epsilon^{2}\geq h^{2}/32\pi^{2}\langle  m_{0}\vert L^{2}\vert m_{0}\rangle$, where $L^{2}$ is the square of the angular momentum operator of the apparatus: to make the error extremely small one has to make extremely large $\langle  m_{0}\vert L^{2}\vert m_{0}\rangle$ .

\subsection{More realistic formalizations of the measurement process}

Up to this point, when accounting for the occurrence of measurement processes, we have always made reference to the projection operators on the eigenmanifolds of the operators associated to the measurement. However, in practice, it is quite difficult to have apparatuses whose effect on the statevector can be accounted precisely by a projection operator. A simple example is the one of  a detector of the position of a particle in a given interval $\Delta$ which has different efficiency in different portions of the interval $\Delta$ so that it detects for sure a particle impinging on  its  central region but only with a certain probability  a particle which is detected near its extreme points. Another example is given by a measurement process which corresponds to two different successive measurements of two noncommuting observables, the outcome being represented by the pair of results which have been obtained. Also in this case the probability of ``an outcome" cannot be expressed in terms of a single projection operator. The appropriate consideration of situations like those just mentioned has led to the consideration of  more general processes affecting the statistical operator than the one of Eq.(3). By taking advantage of a fundamental theorem by Kraus [16] asserting that the most general map of trace class and trace one semipositive definite operators onto themselves  which respects also the condition of complete positivity (which has strong physical reasons to be imposed\footnote{For a definition and a discussion of completely positive maps we refer the reader to ref.[16]}) has the form:
\begin{equation}
\rho\rightarrow\sum_{i} A_{i}^{\dag}\rho A_{i}, \; \sum_{i}A_{i}A_{i}^{\dag}=I.
\end{equation}

When considering measurement processes we will make reference to Eq.(3) or to the just written equation as expressing the effect of the measurement  on the statistical operator.

\section{Proposals of faster-than-light communication and their rebuttal}

As already anticipated, after the clear cut proof by J.S. Bell of the fundamentally nonlocal nature of physical processes involving far away constituents in an entangled state, many proposals have been put forward, either in private correspondence or in scientific papers, suggesting how to put into evidence superluminal effects. We will begin by reviewing a series of proposal whose rebuttal did require only to resort to the standard formalism or to well established facts, such as those put into evidence by the Wigner-Araki-Yanase theorems.

\subsection{Proposals taking advantage of the conservation of angular momentum}

In the year 1979 various papers appeared asserting the possibility of superluminal communication by taking advantage of the change in the angular momentum of a far away constituent due to a measurement performed on its partner. The scientific and social context of these first investigations aiming to take advantage of quantum nonlocality have been described in the interesting and funny book [17] by D. Kaiser {\it How the hippies saved physics},  which intends to point out how the actions of a peculiar community of scientists and non scientists trying to justify various sort of paranormal effects on the basis of quantum nonlocality have drawn, in the US, the attention of the scientific community  to Bell's fundamental theorem and its implications. The three papers [18-20] that I intend to consider in this section have some strict links with the just mentioned context.

Let us  start with refs.[18,19]. Their argument is quite straightforward: one considers two far away spin $1/2$ particles in the singlet state which interact with 2 apparatuses aimed to measure the spin $z$-component  and are in their ``ready" states $\vert A_{0}\rangle$ and $\vert B_{0}\rangle$, so that the initial state is:
\begin{equation}
\vert\Psi\rangle=\frac{1}{\sqrt{2}}[\vert 1_{+},2_{-}\rangle-\vert 1_{-},2_{+}\rangle]\otimes \vert A_{0},B_{0}\rangle.
\end{equation}
Here the indices + and - denote the values (in the usual units) of the $z$-component of the spin of the $i-th$ particle.
Suppose now that the interaction of particle 2 with the apparatus $B$ takes place  before the other particle reaches $A$ ($A$ and $B$ being at rest in a given inertial frame). Wave packet reduction occurs, and we are left, with the same probability, with one of the two states $\vert 1_{+},2_{-},A_{0},B_{-}\rangle$  and  $\vert 1_{-},2_{+},A_{0},B_{+}\rangle$, where $|B_{\pm}\rangle$ are the states of the apparatus $B$ after the measurement. We can now evaluate the mean value of the square of the spin angular momentum when the state is the one of Eq.(11) and when it is one of the states of the mixture. In the first case we get: $\langle S^{2}\rangle_{singlet}=0$, while in the second case we get the value $\hbar ^{2}$. Now one takes advantage of the conservation of the angular momentum ${\bf L=M+S}$ where ${\bf M}$ is the angular momentum of the apparatus\footnote{Here, we disregard the orbital angular momentum of the particles, but the argument holds true also without this limitation.}. Since $\langle{\bf M\cdot S}\rangle=0$ in all above states, one concludes that the measurement induces a change of $\hbar^{2}$ in the apparatus which performs  the measurement. This is not the whole story. If one, subsequently, leaves the second particle to interact with the  apparatus measuring the spin state of the particle, the expectation value of ${\bf L}^{2}$ does not change any more. So, actually, the  angular momentum of the apparatus which is the first to perform the measurement changes of the indicated amount, while the one of the other remains unchanged. Now if Alice and Bob, sitting near $A$ and $B$, have at their disposals a source of entangled particles in the singlet state, Bob, which interacts first with his particle, can choose to perform or not the measurement; correspondingly he can choose whether to leave unchanged or to change the angular momentum of the apparatus at $A$. If Alice can detect this change she can get information about the choice made, for each single instance, by Bob\footnote{It is interesting to remark that the same argument can be developed if one does not take into account the reduction process, i.e., if one assumes that the interactions simply take place in accordance with the von Neumann scheme.}. Superluminal communication becomes possible.

According to the above analysis and the remark in the footnote, the key ingredient which allows to draw the conclusion is the occurrence of an ideal nondistorting measurement of the spin component. This implies that the argument of refs.[18,19] is based on contradictory assumptions, since, as discussed in the previous section, the Wigner-Araki-Yanase theorem asserts that the occurrence of an ideal nondistorting measurement of a spin component  of a subsystem contradicts the conservation of total angular momentum. Actually, to have an ideal measurement process one needs apparatuses with a divergent mean value of the square of the angular momentum, but then no change of this quantity can be detected. Alternatively, one should consider nonideal measurements which are compatible with angular momentum conservation, but then the previous argument does not work, just because Eq.(6) has to be modified.

Precisely in the same year in which the above described arguments were presented, N. Herbert circulated a paper [20] which made resort to the functioning of a half wave plate to get the same result. His proposal was stimulated by his reading of a paper [21] written in 1936 by R. Beth and included by the American Association of Physics Teachers in a collection of papers published as {\it Quantum and Statistical Aspects of Light}. Beth managed to measure the angular momentum of circularly polarized light due to the fact that when right-circularly polarized light is shone on the half-wave plate it sets the plate spinning in one direction, while left-circularly polarized light spun the half-wave plate in the opposite direction. Moreover, the plate flips the light polarization from left to right and viceversa. Beth had measured the effect for circularly polarized light waves, i.e., by using a huge collection of photons all acting together. Herbert, inspired by this work, suggested, in the paper he called QUICK, to play a similar game with the angular momentum of individual photons to get superluminal effects.

Once more the idea is quite simple: one imagines a source emitting pairs of entangled photons in two opposite directions, their state being the analogous of the singlet state, i.e. the rotationally invariant state: $|\Psi (1,2)\rangle=[1/\sqrt{2}][\vert H1,H2\rangle +\vert V1,V2\rangle]\equiv [1/\sqrt{2}][\vert R1,L2\rangle +\vert L1,R2\rangle]$. Here, the symbols $H,V,R$ and $L$ make reference to the states of horizontal, vertical, right and left circular polarizations, respectively. Bob can freely choose whether to perform a measurement of either plane (H,V) or circular (L,R) polarization. As a consequence of his measurement the far away photon is projected either onto a state of plane or of circular polarization. Subsequently, this photon impinges on a half-wave plate (near Alice). Since the photon when plane polarized crosses the plate without transmitting any angular momentum to it, while, when circularly polarized, it imparts a change of $\pm 2\hbar$ to the angular momentum of  the plate, if Alice is able to check whether his plate has not changed or has changed its angular momentum she can know what kind of measurement (H,V) or (L,R) Bob has chosen to perform in any single case. Once more, entanglement and reduction of the wave packet allow superluminal transmission of information.

To prove why also this suggestion is inviable one has to analyze the functioning of the half-wave plate. The nice fact is that, as proved in ref.[22], one can develope precisely an argument  analogous to the one of Wigner-Araki-Yanase, to prove that a half-wave plate can work as indicated only if a violation of the angular momentum conservation occurs. But such a conservation is necessary for the argument, so, once more, the proposal is contradictory. No superluminal communication is possible by resorting to the QUICK device.

Herbert and the Fundamental Fysiks Group made all they could do to spread out Herbert's conclusions. The debate involved scientists like H. Stapp and P. Eberhard. In june 1979, Stapp  challenged the idea, building on Eberhard's argument that statistical averages would wash out any non local effect. But Herbert had worked out his reasoning for individual photons, and the above objection turned out to be not relevant for setting the issue. In the same year we (T. Weber and myself), became acquainted with Herbert's, as well as with Selleri's and others proposals. Accordingly, we  wrote the paper [22] which presents the conclusions I have just described concerning refs.[18-20]. Beth's important experiment worked just because the experimenter sent an enormous number of photons at the half-wave plate. But, at the single-photon level, to get the same result, the half-wave plate would have to be infinitely massive, and, as such, it could not be put into rotation by the passage of an individual photon. This conclusion can be made rigorous  with a little of mathematics, as we did in ref.[22].

\subsection {Popper enters the game} 

Mention should also be made of the position of K. Popper concerning the problem of faster-than-light communication. In some previous writings, but specifically in his famous book  [23] {\it Quantum Theory and the Schism in Physics} he raised the question of the conflict between quantum theory and  special relativity theory, due to his alleged claim that ``if quantum mechanical predictions are correct", then one would be able to send superluminal signals putting into evidence a conflict between the two pillars of our conception of the world. Unfortunately he was (mistankingly) convinced that the quantum rules would imply an effect that they actually exclude (a fact which he missed completely to understand), and, consequently, in his opinion they would allow superluminal signaling in an appropriate experimental situation. 

The idea is quite simple (see fig 1, a,b): we have two perfectly correlated (in position)  particles propagating  towards two arrays of detectors placed at left ($L$) and right ($R$) of the emitting source at almost equal distances from it. Two slits, orthogonal to the  direction $x$ in the figure, are placed at both sides, along the $y$-axis, before the array of the counters, and, initially,  only the counters lying behind the opening of the slits get activated. Subsequently, the slit at $R$ is narrowed so as to produce an uncertainty principle scatter of the momentum $p_{y}$, which appreciably increases the set of counters which are activated with a non-negligible probability (see Fig. 1b). Popper then argues: If quantum mechanics is correct, any increase in the  knowledge of the position $y$ at $R$ like the one we get by making more precise the location along the $y$-axis of the particle which is there, implies an analogous increase of  the knowledge of the position of the particle at $L$. As a consequence also the scatter at $L$ should increase even though the width of the slit at this side has not been narrowed. This prediction is testable, since new counters would be activated with an appreciable probability, giving rise to  a superluminal influence: Alice can know (with an appreciable probability) whether Bob has chosen to narrow or leave unchanged his slit. The conclusion of Popper is quite emblematic: in his opinion the increase of the spread at $L$ would not occur and this {\it would show that quantum theory is wrong}. He also contemplates the other alternative: if the scatter at left would increase, then superluminal communication would be possible and relativity theory would be proven false; in both cases, a quite astonishing conclusion.

\begin{figure}[htb]
\centering
\includegraphics[width=8cm]{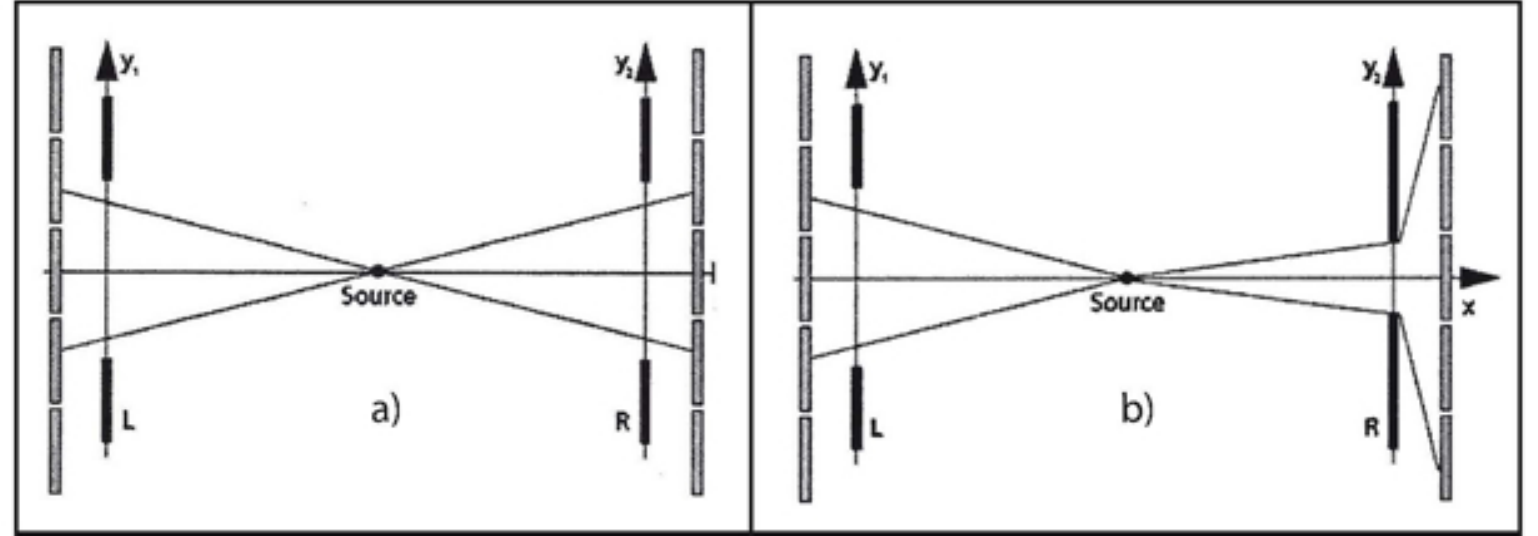}
\caption{The set up and functioning of Popper's ideal experiment.}
\end{figure}

Unfortunately, in this passage of his important work, Popper shows his lack of understanding the quantum principles governing the unfolding the considered experiment. In fact, it can be easily proved that quantum mechanics  predicts precisely that no scatter at left will be induced by the narrowing of the slit at right.  We do not consider it useful to enter in all technical details of the argument. The reader can look at ref.[24] or to Chapter 11 of ref.[25], for a detailed and punctual discussion. Here, we simply outline the argument: if the positions of the particles are really 100\% correlated (and therefore associated to a Dirac delta like unnormalized state), then they are in a state which implies that, even when the two slits are fully opened, all counters are activated with large probabilities, while, if their correlations are only approximate (even with an extremely high degree of accuracy)  the action at $R$ by Bob does not change in any way whatsoever the outcomes at $L$. So, the argument is basically wrong. 

\section{The general proof of the impossibility of faster-than-light communication}

To present a completely general proof [26,27] of the fact that instantaneous wave-packet reduction does not allow superluminal signaling we must start be reconsidering all possible actions [3,16] which are permitted, by standard quantum mechanics, on a constituent of a composite system. Quite in general, quantum mechanics allows the possibility of:
\begin{itemize}
\item A unitary transformation describing the free evolution of the system at $R$ under consideration and/or possibly of its interactions with other systems lying in a space-time region which is space like with respect to the one at $L$. 
\item A transformation corresponding to a non selective projective measurement (with wave packet reduction as described by Eq. (3)) of the considered subsystem.
\item A transformation  like the one summarized in Eq.(10) corresponding, essentially, to the occurrence of a non ideal measurement.
\item A  transformation like the one of Eq.(4), corresponding to a selective measurement. To be strict, one should also consider the analogous of this transformation in the case of a non-ideal measurement, but, for the reasons which we will make clear below, this case does not have a physical relevance for faster-than-light signaling. 
\end{itemize}
  
Now we can proceed to outline our proof, which represents a generalization and a more accurate formulation of some ideas put forward by P. Eberhard [28] about one year before we wrote our paper [26] (see also A. Shimony [29] for an enlightening discussion).

To start with we recall  that all probabilistic  predictions concerning a subsystem of a composite system can be obtained by considering the reduced statistical operator of the subsystem of interest. We suppose now to have a composite quantum system $S=S^{(1)}+S^{(2)}$ associated to the statistical operator $\rho(1,2)$ and to be interested in predictions concerning prospective measurements on subsystem $S^{(1)}$. As already remarked, the physics of this subsystem is fully described by the reduced statistical operator $\tilde{\rho}^{(1)}=Tr^{(2)}\rho(1,2)$. We can now consider the following set of equations:
\begin{eqnarray}
\tilde{\rho}^{(1)} &=&Tr^{(2)} \rho(1,2),\nonumber \\
Tr^{(2)}[U^{(\dag)(2)}\rho(1,2)U^{(2)}]&=&Tr^{(2)}[U^{(2)}U^{(\dag(2)}\rho(1,2)]=Tr^{(2)} \rho(1,2)=\tilde{\rho}^{(1)},\nonumber\\
Tr^{(2))}[\sum_{i} P^{(2)}_{i}\rho(1,2)P^{(2)}_{i}]&=&Tr^{(2)}[\sum_{i} [P_{i}^{(2)}]^{2}\rho(1,2)]=Tr^{(2)}\rho(1,2)=\tilde{\rho}^{(1)}\nonumber \\
Tr^{(2))}[\sum_{i} A^{\dag(2)}_{i}\rho(1,2)A^{(2)}_{i}]&=&Tr^{(2)}[\sum_{i} A_{i}^{(2)}A^{\dag(2)}_{i}\rho(1,2)]=Tr^{(2)}\rho(1,2)=\tilde{\rho}^{(1)}. 
\end{eqnarray}

In these equations we have made use of the cyclic property of the trace over the Hilbert space $\cal{H}$$^{(2)}$ when operators of the same Hilbert space are involved, of the unitarity relation $U^{(2)}U^{(\dag)(2)}=I^{(2)}$, of the propery $[P^{(2)}_{i}]^{2}=P^{(2)}_{i}$ and of the fact that the sum of the operators $P^{(2)}_{i}$ in the case of a projective measurement, as well as of the operators $A_{i}^{(2)}A^{\dag(2)}_{i}$ of the Kraus theorem [16], must equal the identity operator of the same space. We call the attention of the reader to the fact that in all considered cases, i.e., i). no action on $S^{(2)}$, ii). a unitary evolution of  $S^{(2)}$, or iii). the fact that it is subjected to an  operation corresponding to a (ideal or non-ideal) nonselective measurement process,  the reduced statistical operator $\tilde{\rho}^{(1)}$ of subsystem $S^{(1)}$ does not change in any way whatsoever, and, accordingly, Alice,  performing measurements on such a subsystem cannot get any information about the fact that Bob is making some specific action on subsystem $S^{(2)}$.

Up to now, we  have not considered explicitly the case of selective ideal or non-ideal measurement processes, accounted for by Eq. (4) or by its analogue referring to processes like those governed  by Eq.(10). If one considers the modifications to the general statistical operator in these cases and one uses the reduced statistical operator to evaluate the probabilities of the measurement outcomes on subsystem $S^{(1)}$, one would easily discover that the physics of such a system is actually changed by the action on its far-away partner. But the probabilistic changes depend crucially on the outcome that Bob has obtained in his measurement, so that Alice might take advantage of this fact only if she would be informed of the outcome obtained by Bob. This implies that Bob must inform Alice concerning his outcome  and this can be done only by resorting to standard communication procedures which require a subluminal communication between the two. Accordingly, these cases can safely be disregarded within our context. 

It should be clear that the general validity of our theorem implies that all previously discussed attempts to get faster-than-light signaling taking advantage of the instantaneous reduction at-a-distance of the statevector in the case of entangled states of far-away systems, were doomed to fail. We have discussed them in some detail to present an historically complete perspective of the debate on this fundamental issue, i.e. the one of the compatibility of quantum mechanics with relativistic requirements concerning the communication between far-away observers.

\section{A radical change of perspective}
\subsection{Herbert's new proposal}
In  1981 N. Herbert submitted for publication to Foundations of Physics a paper [4] by the title: {\it FLASH--A superluminal communicator based on a new kind of quantum measurement} in which he  added a new specific device to his previous proposal we have discussed in sect.3.1. The stimulus to do so came probably from our paper, ref.[26], as remarked by D. Kaiser in his book [17]: {\it From Ghirardi's intervention, Herbert came to appreciate the importance of amplifying the tiny distinction between various quantum states, to evade fundamental limits on signaling.} The crucial device which, in his opinion, could do the game, was a Laser gain tube exhibiting the following characteristics: if the laser was stimulated by a single photon in any state of polarization (the states which mattered were actually those of plane (V and H) and of circular (R and L) polarization) it would emit a relevant number, let us say 4N with N large, of identical copies (in particular with the same polarization) of the impinging photon.  If we summarize the process by means of an arrow leading from the initial photon state to the bunch of final photons, Hebert's Laser gain tube has to work in the following way:
\begin{equation}
\vert V,1\rangle\rightarrow\vert V,4N\rangle,\;\vert H,1\rangle\rightarrow\vert  H,4N\rangle,\;\vert R,1\rangle\rightarrow\vert R,4N\rangle,\;\vert L,1 \rangle\rightarrow\vert L,4N\rangle,
\end{equation}

with obvious meaning of the symbols. Here 1 denotes the photon propagating towards Alice.
 
By resorting to this machine  Herbert's game  becomes quite simple. One starts, as in his first proposal, with a source  emitting pairs of entangled photons in two opposite directions, their state being  the rotationally invariant state: $|\Psi (1,2)\rangle=[1/\sqrt{2}][\vert H1,H2\rangle +\vert V1,V2\rangle]\equiv[1/\sqrt{2}][\vert R1,L2\rangle +\vert L1,R2\rangle]$.  Obviously,  Bob can freely choose whether to perform a measurement of either plane (H,V) or circular (L,R) polarization. As a consequence of his measurement the far away photon is projected either onto a state of plane or of circular polarization. At this point the far away photon is injected in the Laser gain tube which emits $4N$ photons with the same polarizations, which, in turn, depends on the measurement performed by Bob and the outcome  he has got. The $4N$ photons are then separated into 4 beams of N photons each, directed towards 4 detectors of $V$,$H$,$L$ and $R$ (mind the order) polarization, respectively. To see the game coming at an end we have now simply to recall that a detector registers for sure a photon with the polarization it is devised to measure, it does not detect a photon into an orthogonal state and detects with probability $1/2$ a photon in a state of polarization which is the equal superposition of the state that it is devised to detect and of the state orthogonal to it.

We analyze  the situation in detail specifying the measurements which Bob chooses to perform, the outcomes he gets and the records by the counters near Alice.

\begin{itemize}
\item Suppose Bob choses to perform a polarization measurement aimed to ascertain whether the photon (2) reaching him has vertical or horizontal polarization and that he finds the photon vertically polarized. In this case, the process goes as follows:

a). Initial state: $|\Psi (1,2)\rangle=[1/\sqrt{2}][\vert H1,H2\rangle +\vert V1,V2\rangle]$; b). Measurement with outcome Vertical; c). Reduction of the state: $\vert V1, V2\rangle$; d). Amplification: $\vert V,4N;V2\rangle$; e). Number of photons detected by the far away detectors (near Alice) for the 4 beams: N,0,N/2,N/2.

\item 
a). Initial state: $|\Psi (1,2)\rangle=[1/\sqrt{2}][\vert H1,H2\rangle +\vert V1,V2\rangle]$; b). Measurement with outcome Horizontal; c). Reduction of the state: $\vert H1,H2\rangle$; d). Amplification: $\vert H,4N;H2\rangle$; e). Number of photons detected by the far away detectors on the 4 beams: 0,N,N/2,N/2.

\item 
a). Initial state: $[1/\sqrt{2}][\vert R1,L2\rangle +\vert L1,R2\rangle]$; b). Measurement with outcome Right; c). Reduction of the state: $\vert L1, R2\rangle$; d). Amplification: $\vert L,4N;R2\rangle$; e). Number of photons detected by the far away detectors on the 4 beams: N/2,N/2,N,0.
\item 
a). Initial state: $[1/\sqrt{2}][\vert R1,L2\rangle +\vert L1,R2\rangle]$; b). Measurement with outcome Left; c). Reduction of the state: $\vert R1; L2\rangle$; d). Amplification: $\vert R,4N;L2\rangle$; e). Number of photons detected by the far away detectors on the 4 beams: N/2,N/2,0,N.
\end{itemize}

Now, one has simply to remark that in the cases listed under the two first items (i.e. when Bob chooses to measure linear polarization) the detector which does not register any photon is either the first or the second, while, in the alternative case in which Bob chooses to measure circular polarization, it is either the third or the fourth detector which does not register any photon. Accordingly, Alice can become aware, instantaneously, of the choice made by Bob:  they can communicate superluminally.

\subsection{The no-cloning theorem}

The FLASH paper was sent for refereeing to A. Peres and to me.  Peres' answer [30] was rather peculiar: {\it I recommended to the editor that this paper should be published. I wrote that it was obviously wrong, but I expected that it would elicit considerable interest and that finding the error would lead to significant progress in our understanding of physics}. I also was  rather worried for various reasons.   I was not an expert on Lasers and I was informed that A. Gozzini and R. Peierls were trying to disprove Herbert's conclusion by invoking the unavoidable noise affecting the Laser which would inhibit its desired functioning. On the other hand, I was convinced that quantum theory in its general formulation and not due to  limitations of practical nature would make unviable Herbert's proposal.  After worrying for some days about this problem I got the general answer: while it is possible to devise an ideal apparatus which clones two orthogonal states  with 100\% efficiency, the same apparatus, if the linear quantum theory governs its functioning, cannot clone states which are linear combination of the previous ones. Here is my argument, on the basis of which I recommended rejection of Herbert's paper. The assumption that  the cloning machine acts as follows:
\begin{equation}
\vert V\rangle\rightarrow\vert V,4N\rangle \;\;and\;\;\vert H\rangle\rightarrow\vert H,4N\rangle, 
\end{equation}

when the linear nature of the theory is taken into account, implies:
\begin{equation}
\vert R\rangle\equiv\frac{1}{\sqrt{2}}[\vert V\rangle+\vert H\rangle]\rightarrow\frac{1}{\sqrt{2}}[\vert V,4N\rangle+\vert H,4N\rangle],
\end{equation}

and analogously for the left polarization. Now,  the state at the r.h.s. of the last equation is by no means the state $\vert R,4N\rangle$ which Herbert had assumed to occur in the case in which the Laser gain tube is triggered by a right polarized photon. But this is not the whole story: how it has been shown in ref.[31] the very linear nature of the theory implies that no difference in the detections of Alice occurs in dependence of the free will choice of Bob.

This is an account of how Herbert's ingenious, but mistaken, proposal has led me to be the first to derive the no-cloning theorem \footnote{I have chosen to attach at the end of the paper, a document - a letter by A. van der Merwe - which officially attests this fact, since it is known only to a restricted community of physicists.}.  About one and half  year later  Wootters and Zurek [32] and  Dieks [33] derived independently the same result and published it\footnote{I must confess that I have never understood why A. Peres, in mentioning my derivation, has stated that it was a special case of the theorems in refs. [32] and [33]. Comparison even only of the short page by A. van der Merwe with the just mentioned papers makes clear that the argument is precisely the same and has the same generality.}. The theorem is of remarkable importance in quantum theory, it has become known as ``The no cloning theorem" and  it has been quoted an innumerable number of times. Only subsequently I realized that it had been a mistake on my part not to publish my result. I discussed my precise argument with Gozzini and Peierls, by sending them a draft which was a sort of repetition of my referee report and I subsequently published it [31] in collaboration with my collaborator, T. Weber.

\subsubsection{More on quantum cloning}

In a paper like the present one, we believe it useful to  mention that  E.P. Wigner [34], in an essay of 1961  had already argued that the phenomena of self-replication of biological systems contradicted the principles of quantum mechanics. His argument is quite straightforward. Following his notation let us suppose we have a living system in a state $\nu$ and an environment (assimilated to  ``food") in a state $w$, so that, the initial statevector of the system, organism + nutrient, is:
$\Phi=\nu \times w$. When replication takes place the statevector will have the form: $\Psi=\nu \times \nu \times r$, i.e. two organisms each in the statevector $\nu$ will be present, while the vector $r$ describes both the rest of the system,  the rejected part of the nutrient and also the other coordinates (positions, etc.) of the two organisms. One assumes that the system lives in an $N$-dimensional Hilbert space $\cal{H}^{(N)}$, the part $r$ in an $R$-dimensional Hilbert space $\cal{H}^{(R)}$, while the ``food" state $w$ belongs to a $N\cdot R$ dimensional Hilbert space $\cal{H}^{(NR)}$, so that $\Phi$ and $\Psi$ live, as they must, in the same space. Suppose we do not know the state of the living system; however, since it belongs to  $\cal{H}^{(N)}$ his knowledge requires to know $N$ complex numbers. Analogously we do not know the state $r$, and the state $w$, which require the specification of $R$ and $NR$ complex numbers to be determined. We now assume, with Wigner. that the collision matrix which gives the final state resulting from the interaction, which will be denoted as $S$, of the organism and the nutrient is a random matrix, which, however, even though unknown to us explicitly, is completely determined by the laws of quantum mechanics. Obviously $S$ must satisfy:
\begin{equation}
\Psi=S\Phi.
\end{equation}
Choosing the direct product of a basis $\{ \nu_{k}\}$ for  $\cal{H}^{(N)}$, one  $\{r_{\mu}\}$ for  $\cal{H}^{(R)}$  and one $\{w_{\lambda,\mu}\}$ for $\cal{H}^{(RN)}$, and projecting Eq. (16) on such a basis one gets   $N^{2}R$  equations. And now the conclusion follows: our unknown quantities are the components of the states $\nu, r$ and $w$ on their respective bases and are therefore $N+R+NR$ in number. Thus, according to Wigner, the question is: given the matrix corresponding to $S$, it is possible to find vectors $\nu$,$r$ and $w$ such that their components satisfy the above mentioned $N^{2}R$ equations? Since $N^{2}R \gg N+R+NR$, for extremely large $N$ and $R$, according to him: {\it it would be a miracle if such equations could be satisfied}. In other words, a self-replicating quantum unit does not exist. One might state that Wigner has ``derived" (with the proviso he is making - see below) the no-cloning theorem for a quantum system whose Hilbert space has an extremely high dimensionality N, while we have shown that it holds also for N=2.

Wigner was perfectly aware that the argument is not fully rigorous and cannot be taken too seriously because of the many assumptions on which it is based. However,  he seems inclined to attach a certain value to it. This is not surprising because at the time in which he wrote his paper he was adhering to von Neumann's idea that consciousness is responsible for the reduction of the wave packet, so that, in a certain sense, the fact that quantum mechanics is not able to account for the basic property of living organism (the self-reproduction) supported his view that such a theory cannot be used to describe the conscious perceptions of such organisms. In 1971 Eigen [36] responded to Wigner claiming that his choice  of resorting to a typical unitary map to account for the process did not take into account the instructive functions of informational macromolecules.

Strictly connected with Wigner argument, even though derived through a much more rigorous and general procedure is the proof of the no-cloning theorem presented in a beautiful paper by R. Alicki [35]. He considers a dynamical transformation T from an initial state $\varphi\otimes\omega$, where $\varphi$ is the state of the organism and $\omega$ the fixed initial state of the environment  designed as ``food":
\begin{equation}
\varphi\otimes\omega\rightarrow T(\varphi\otimes\omega)=\varphi\otimes\varphi\otimes\sigma.
\end{equation}
As before, $\sigma$ represents the state of the ``food" after the replication. Alicki assumes that any dynamical process of a closed system (typically the one given by $T(\varphi)$) cannot reduce the  indistinguishability of two states $\varphi$ and $\psi$, which can be quantified by the ``overlap" $(\varphi|\psi)$ of the two states\footnote{Here , the expressions $(\alpha|\beta)$ must not be  identified with the Hilbert scalar product, since Alicki is taking a much more general perspective, which, however, requires to quantify the idea of distinguishability and its fundamental properties. He does this by introducing his symbol for the overlap.}, i.e., $(T(\varphi)|T(\psi))\geq (\varphi |\psi)$ (which  in Alicki's spirit can be considered as a form of the second law of thermodynamics: indistinguishability cannot decrease with the evolution). Then one has:
\begin{eqnarray}
(\varphi|\varphi')&=&(\varphi\otimes\omega|\varphi'\otimes\omega)\leq(T(\varphi\otimes\omega)|T(\varphi'\otimes\omega))\nonumber \\
&=& (\varphi\otimes\varphi\otimes\sigma|\varphi'\otimes\varphi'\otimes\sigma')=(\varphi|\varphi')^{2}(\sigma|\sigma')\leq (\varphi|\varphi')^{2},
\end{eqnarray}

implying $(\varphi|\varphi')=1$ or $(\varphi|\varphi')=0$. 
It is interesting to note that if, taking a strictly quantum perspective (which means to replace the round brackets in the above equation by Dirac's bras and kets), one identifies (as it is quite reasonable) the general concept of overlap$\equiv$indistinguishability with the scalar product of the Hilbert space and one assumes that the unfolding of the process is governed by a unitary transformation (which as such does not change the overlap), the above proof corresponds to a modern version of the no-cloning theorem which, in place of using  the linearity of the evolution as we and the authors of refs.[32,33] did in deriving the theorem, makes resort to unitarity. 

\section{Further recent proposals which require new impossibility proofs}

\subsection{A proposal by D. Greenberger}

In spite of the lively debate and the many precise results which should have made fully clear why quantum mechanics does not allow superluminal communication, new papers claiming to have found a new way to achieve this result continue to appear. The first we want to mention is a proposal [37] of D. Greenberger which has been considered as inspiring even quite recently. Actually, in ref.[38] it is claimed that the proposal of Greenberger {\it has not yet been refused and calls into question the universality of the no-signaling theorem}, and, accordingly, it represents a stimulus to pursue the investigations on this line. 

Greenberger proposal involves the simoultaneous emission of two photons  by a source along two different opposite directions $(a,a')$ and $(b,b')$, so that the initial state is the entangled state:
\begin{equation}
\vert\psi\rangle_{1,2}=\frac{1}{\sqrt{2}}[\vert a\rangle_{1}\otimes \vert a'\rangle_{2}+\vert b\rangle_{1}\otimes \vert b'\rangle_{2}]
\end{equation}

\noindent Subsequently, the two photons  impinge on a series of beam splitters, as shown in figure 2. The horizontal gray boxes represent the beam splitters which are assumed to both reflect and transmit half the incident light, and produce a phase shift of $\pi/2$ upon reflection and none upon transmission. On the path of the photon emitted along $b$, after it goes through the first beam splitter, there is a phase-shifter $A$ that shifts the phase of any photon passing through it by $\pi$, and that can be inserted or removed from the beam at will.
\begin{figure}[htb]
\centering
\includegraphics[width=8cm]{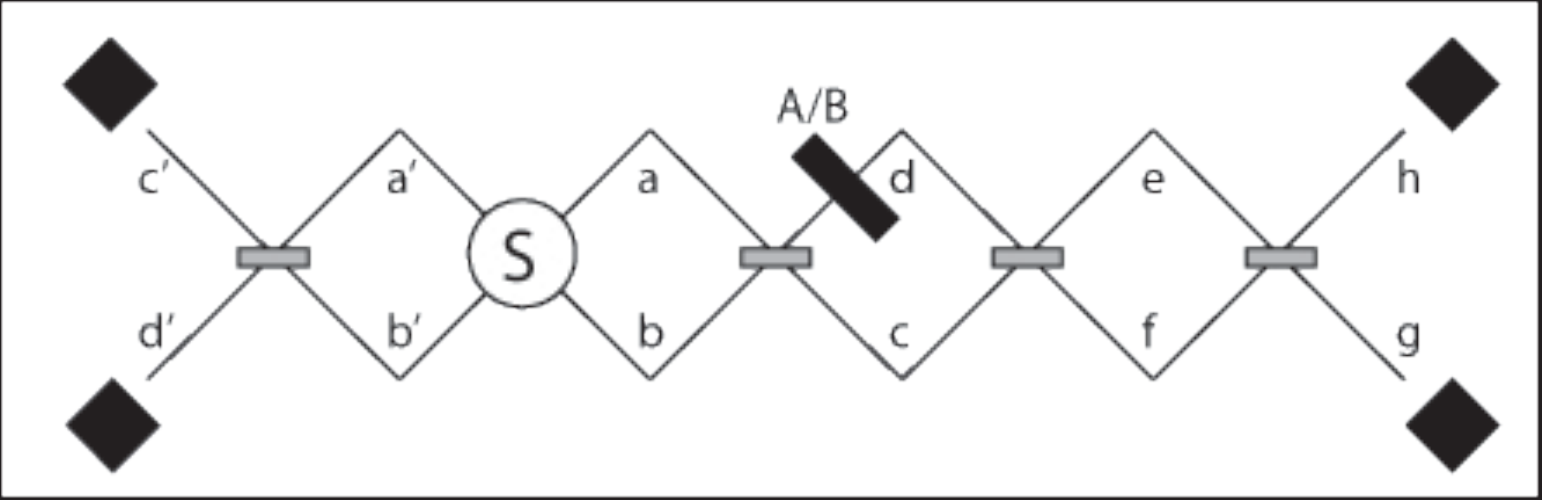}
\caption{Illustration of Greenberger's proposal as depicted in his paper.}
\end{figure}

At this point the first crucial assumption  of the paper enters into play:
\begin{quote}
 i). {\it The phase shifter can be prepared not only in the states $\vert A\rangle$ and $\vert B\rangle$, corresponding to its being inserted or removed from the path of the photon, but also in their orthogonal linear combinations:}
\begin{equation}
\vert u\rangle_3 = \frac{1}{\sqrt{2}}[\vert A\rangle + \vert B\rangle], \qquad
\vert v\rangle_3 = \frac{1}{\sqrt{2}}[\vert A\rangle - \vert B\rangle].
\end{equation}
\end{quote}
According to the author of [37], one can also switch on the Hamiltonian $H$ for this macroscopic object,
whose eigenstates are $\vert u \rangle_3$ and $\vert v \rangle_3$, corresponding to slightly different energies, implying the development in time of relative phases with respect to each other.

We will not go through the subsequent elementary calculations of the paper; we limit ourselves to mention that the above assumptions lead to the conclusion that, as the photons are nearing their final detectors represented in the figure by the $4$ black squares, they will be in the following entangled photon-phase shifter state:
\begin{equation}
\vert \psi \rangle_{1,2,3} = \frac{1}{2}\big[(-e^{i\alpha}\vert h\rangle_{1}\vert d'\rangle_{2}+e^{-i\alpha}\vert g\rangle_{1}\vert c'\rangle_{2})e^{i\beta}\vert u\rangle_3
+ (e^{i\alpha}\vert g\rangle_{1}\vert c'\rangle_{2}-e^{-i\alpha}\vert h\rangle_{1}\vert d'\rangle_{2})e^{-i\beta}\vert v\rangle_3\big],
\end{equation}
the phase factors $e^{\pm i\alpha}$ and $e^{\pm i\beta}$ being due to the evolution of the states $\vert u\rangle_3$ and $\vert v\rangle_3$.

At this point Greeneberger  puts forward his really crucial assumption. In his words:
\begin{quote}
ii). {\it  In accordance with our assumption that one can manipulate these Cat states, one can turn off $H$ for the state $\vert v\rangle$, while leaving it in place for the state $\vert u\rangle$. This will rotate the state $\vert v\rangle$ into the state $e^{i\gamma}\vert u\rangle$, where $\gamma$ is the accumulated phase difference during this process.}
\end{quote}

As it is obvious this amounts to accept that a nonunitary transformation $T$ can be performed:
\begin{equation}\label{madtransf}
T\vert u\rangle_3 =\vert u\rangle_3,\qquad T\vert v\rangle_3=e^{i\gamma}\vert u\rangle_3.
\end{equation}
The conclusion follows. After  this transformation the state becomes:
\begin{equation}
\vert \psi_{final}\rangle_{1,2,3} = e^{i\gamma/2} \big[-\cos (\alpha+\beta-\gamma/2)\vert h\rangle_{1}\vert d'\rangle_{2} + \cos (\beta-\alpha-\gamma/2)\vert g\rangle_{1}\vert c'\rangle_{2}\big]\vert u\rangle_3.
\end{equation}
And now the game is over: by appropriately choosing the angles $\alpha,\beta$ and $\gamma$, one can, at his free will, suppress one of the two terms of the superposition of the photon states, i.e. one can make certain either the firing of the detector in $d'$ or the one in $c'$ (and correspondingly the one in $h$ or the one in $g$) allowing in this way a superluminal transfer of information from the phase shifter, which acts as the signaler, to the photon detectors.

The paper, since  nobody had discussed it in spite of its revolutionary character, deserved some attention;  we have reconsidered it in ref.[39]. Its weak points are:
\begin{itemize}
\item The assumption that one can prepare the linear superposition of two macroscopically different states,  corresponding to different locations of the macroscopic phase-shifter. This is  impossible to get in practice.
\item However, even ignoring the above critical feature of the hypothetical experiment, the really crucial and unacceptable fact is the one embodied in its second assumption, i.e., the possibility of implementing a nonunitary transformation.
\end{itemize}

We will not go on analyzing all the details of ref.[37] and of the punctual criticisms of ref.[39]. We believe that to show where it fails the simplest way is to resort to an example that we have devised in our paper. We consider an elementary EPR-Bohm like setup for two far away spin $1/2$ particles in the singlet state:
\begin{equation}\label{singlet}
\vert \psi_{-}\rangle=\frac{1}{\sqrt{2}}[\vert \uparrow_{1}\rangle \vert \downarrow_{2}\rangle -\vert \downarrow_{1}\rangle \vert \uparrow_{2}\rangle].
\end{equation}
\noindent In strict analogy with what has been assumed by Greenberger, suppose now we can rotate only one of the two spin states of particle $2$ making it to coincide, apart from a controllable phase, with the other one:
\begin{equation}\label{transf}
T\vert \downarrow_{2}\rangle = \vert \downarrow_{2}\rangle, \qquad T \vert \uparrow_{2}\rangle=e^{i\gamma}\vert \downarrow_{2}\rangle.
\end{equation}
Under this transformation the state [25] becomes a factorized state of the two particles:
\begin{equation}\label{final2}
\vert \psi_{T}\rangle = \frac{1}{\sqrt{2}}[\vert \uparrow_{1}\rangle - e^{i\gamma}\vert \downarrow_{1}\rangle] \vert \downarrow_{2}\rangle
\end{equation}
In (26), the state referring to particle $1$ is an eigenstate of $\sigma\cdot {\bf d}$ for the direction ${\bf d}=(\cos \gamma, \sin \gamma, 0)$ pertaining to the eigenvalue $-1$. This means that a measurement of this observable by Alice (where particle 1 is) will give with certainty the outcome $-1$ if Bob has performed the transformation $T$ on his particle, while, if Bob does nothing, the probability of getting such an outcome equals $1/2$. Having such a device, one can easily implement superluminal transfer of information. Concluding: if assumption ii) were correct, one would not need all the complex apparatus involved in Greenberger's proposal which, at any rate, cannot work as indicated due to the nonlinear nature of $T$.

\subsection{A proposal involving a single system}

Another proposal that has to be mentioned is the one [40] by Shiekh. His suggestion is different from all those which have appeared in the literature since the author does not make resort to an entangled state of two systems  but he works with a single particle in a superposition of two states corresponding to its being in two far-away regions, and the measurement process  involves only one of the two far-away parts of the wave function. So, in a sense, the argument of ref.[40] does not fall under the no-go theorems considered here and requires a separate comment. The author is inspired by the fact that when a single particle is associated to a wavefunction as the one just mentioned, any attempt to test whether it is ``here" (at right), or ``there" (at left) changes instantaneously the wavefunction on the whole real axis by making it equal to zero or enhancing it ``there" according to whether we detect or we do not find the particle ``here". The process seems to exhibit some nonlocal aspects due to the instantaneous change at-a-distance. Obviously, that this might lead to superluminal signaling is something that nobody can believe, but it is instructive to show that also in this case, to achieve the desired result, one has to resort  to a nonunitary evolution. The  elementary analysis which follows  will lead once more to the conclusion that the process cannot be used to send superluminal signals. 

We briefly review the argument by Shiekh. He considers a particle which is prepared, at time $t=0$, in an equal weights superposition of two normalized states, $\vert h+\rangle$ and $\vert h-\rangle$, propagating in two opposite directions, respectively, starting from the common origin of the $x$-axis:
\begin{equation}
\vert \psi,0\rangle=\frac{1}{\sqrt{2}}[\vert h+\rangle+\vert h-\rangle].
\end{equation}
Subsequently  the state $\vert h+\rangle$ is injected in an appropriate device behaving in a way rather similar, apart from the final stage, to a Mach-Zender interferometer. One also assumes that an observer, located near to it, can choose, at his free will, to insert or not a phase-shifter along one of the two paths of the interferometer. The two wave functions are then recombined by appropriate deflectors so that, by deciding whether or not to insert the phase-shifter, one can produce a constructive (no phase-shifter in place) or a destructive (the phase-shifter is present) interference of the two terms in which the impinging state  $\vert h+\rangle$ has been split. Finally, a detector is placed along the direction of propagation of the final state and it induces wave packet reduction, since it either detects or fails to detect the particle. We have summarized the  situation for the two considered cases in Figs. 3a,b.
\begin{figure}[htb]
\centering
\includegraphics[width=11cm]{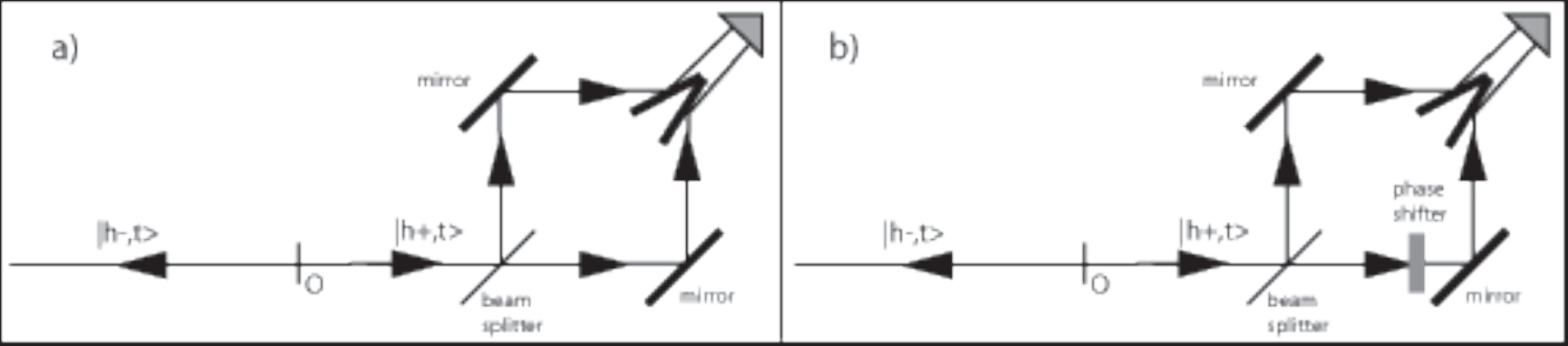}
\caption{The experimental arrangement of ref.[40]. The two cases refer to no Phase-shifter inserted or phase shifter inserted, respectively.}
\end{figure}
The author then concludes: {\it If the sender} (the guy who can choose to insert or not the phase shifter),  {\it arranges for constructive interference, then some of the particles will be "taken up" by the sender, but none if destructive interference is arranged; in this way the sender can control the intensity of the beam detected by the receiver} (the observer located far away where the evolved of $\vert h-\rangle$ is concentrated). {\it So, a faster than light transmitter of information (but not energy or matter) might be possible}.

We believe that all readers will have clear the trivial mistake of the paper. In fact, what one can govern by deciding whether to insert or not the phase-shifter, is the interference at the central region of the final detector. Let us concentrate our considerations only on the normalized state $|h+\rangle$. If triggered by such a state when it exhibits constructive interference, the counter will register (practically) for sure the particle described by such a state (the wavefunction has a peak just there), while, if there is destructive interference, the counter will not register the particle. But this does not mean that the wavefunction associated to $|h+\rangle$ disappears, as claimed by the author, it simply means that its support lies outside the interval covered by the counter. Actually, no one will doubt that if one places an array of counters covering all the line orthogonal to the final direction of propagation, one of them, different from the one of the experiment, will fire for sure. If one combines these considerations with the fact that actually the whole state of the particle is the superposition of $|h-\rangle$  and $|h+\rangle$ one realizes that the statements we have just made  concerning what is going on at right have only probability $1/2$ of occurrence, since the particle can be not detected in the right region. Accordingly, the probability that the particle is found at left remains equal to $1/2$, as if no specific action would be made at right.

It seems rather peculiar that the author introduces an hypothetical process which can make zero a wavefunction (i.e. the normalized state $|h+\rangle$), and as such it does not preserve unitarity, and, at the same time, he resorts to the overall conservation of probability (i.e. to unitarity) to claim that the action at right can change the norm of the state at left.

This concludes our analysis of the many proposals which have been presented to send superluminal signals.

\section{Nonlocality and relativistic requirements}
As already stated, quantum mechanics suffers of an internal inconsistency, the one between the linear and deterministic evolution induced by Schr\"{o}dinger's dynamics and the nonlinear and stochastic collapse of the state in a measurement. Many scientists, among them Einstein, Schr\"{o}dinger, Bell and many others have been disturbed not only by the formal inconsistency between the two dynamical principles, but especially by the fact that the borderline between what is classical and what is quantum, what is reversible and what is irreversible, what is micro and what is macro is to a large extent ambiguous. Accordingly, many serious attempts to overcome this difficulty have been presented, inspired by the conviction that Bell has expressed [6] so lucidly:
\begin{quote}
{\it Either the wavefunction, as given by the Schr\"{o}dinger equation is not everything or it is not right.}
\end{quote} 
\subsection{Bohmian Mechanics}
The first alternative corresponds to the idea that the specification of the state of a physical system given by the statevector has to be enriched or replaced by  new variables (the so-called hidden variables). The best known and rigorous example of this line is represented by Bohmian mechanics [42], a deterministic theory such that the assignement of the wavefunction and of the hidden variables (i.e. the initial positions of all particles which are chosen to be distributed according to the quantum probability $|\psi({\bf r}_{1},{\bf r}_{2},...,{\bf r}_{N}, t_{0})|^{2}$) determines uniquely their positions  at any subsequent time. The predictions of the theory concerning the probability distribution of the particles coincide with those of standard quantum mechanics and the theory overcomes the measurement problem in a clean and logically consistent way. 
 
I will spend only few words on the locality issue within Bohmian mechanics. Since this theory agrees with quantum mechanics in general and typically in an EPR-like situation, it must exhibit a specific sort of nonlocality. It has been proved [43] that any deterministic hidden variable theory equivalent to quantum mechanics admits only relativistic generalizations which must resort to a specific foliation of space-time. In other words, such theories are characterized by a preferred reference frame which, however, remains unaccessible. Accordingly, as stressed by Bell [8], they require a change of attitude concerning Lorentz invariance: the situation resembles the one of the theory of relativity in the Fitzgerald, Larmor, Lorentz and Poincar\'{e} formulation in which there is an absolute aether, but the contraction of space and the dilation of time fooled the moving observers by allowing them to consider themselves at rest. In spite of this remark, explicit and interesting relativistic generalizations of Bohmian mechanics have been presented. In particular  bohmian-like relativistic models have been worked out  both in first quantized versions [44] as well as in the framework of quantum field theories [45].
\subsection{Collapse theories}
The second alternative corresponds to assume that Schr\"{o}dinger's equation has to be changed in such a way not to alter the well established predictions of quantum mechanics for all microscopic systems while leading to the collapse of the statevector with the desired features and probabilities when macroscopic systems enter into play. The first explicit example of this kind is the so called GRW theory [5] which we summarize in a very sketchy way.

The central idea is to modify the linear and deterministic evolution equation of standard quantum mechanics by adding nonlinear and stochastic (i.e. sharing the features of the reduction process) terms to it. As it obvious, and as it has been stressed by many scientists, since the situation characterizing macro-objects correspond to perceptually different locations of their macroscopic parts (e.g. the pointer) the change in the dynamics must strive to make definite the positions of macroscopic bodies. The model is based on the following assumptions:
\begin{itemize}
\item A Hilbert space $\cal{H}$ is associated to any physical system and the state of the system is represented by a normalized vector $|\psi_{t}\rangle$ of $\cal{H}$,
\item The evolution of the system is governed by Schr\"{o}dinger's equation. Moreover, at random times, with mean frequency\footnote{Actually this frequency must be made proportional to the mass of the particles entering into play. The value we will chose below refers to  nucleons.} $\lambda$, each particle of any system is subjected to random spontaneous localization processes as follows. If particle $i$ suffers a localization then the statevector changes according to:
\begin{equation}
|\psi_{t}\rangle\rightarrow\frac{L_{i}({\bf x})|\psi_{t}\rangle}{\parallel L_{i}({\bf x})| \psi_{t}\rangle\parallel}; \;\;L_{i}({\bf x})=(\frac {\alpha}{\pi})^{3/4}e^{[-\frac{\alpha}{2}({\bf \hat{x}}_{i}-{\bf x})^{2}]},
\end{equation}
where ${\bf \hat{x}}_{i}$ is the position operator of particle $i$,
\item The probability density for a collapse at {\bf x} is $p({\bf x})=\parallel L_{i}({\bf x})| \psi_{t}\rangle\parallel^{2}$, so that localizations occur more frequently where the particle has a larger probability of being found in a standard position measurement.
\end{itemize}

The most relevant fact of the process is its ``trigger mechanism", i.e. the fact that, as one can show  by passing to the centre-of-mass and relative coordinates, the localization frequency of the c.o.m. of a composite system is amplified with the number of particles, while the internal motion, with the choice for $\alpha$ we will make, remains practically unaffected. We have summarized the situation for a micro (at left) and macroscopic (at right -- a pointer) system in Fig.4.
\begin{figure}[htb]
\centering
\includegraphics[width=11cm]{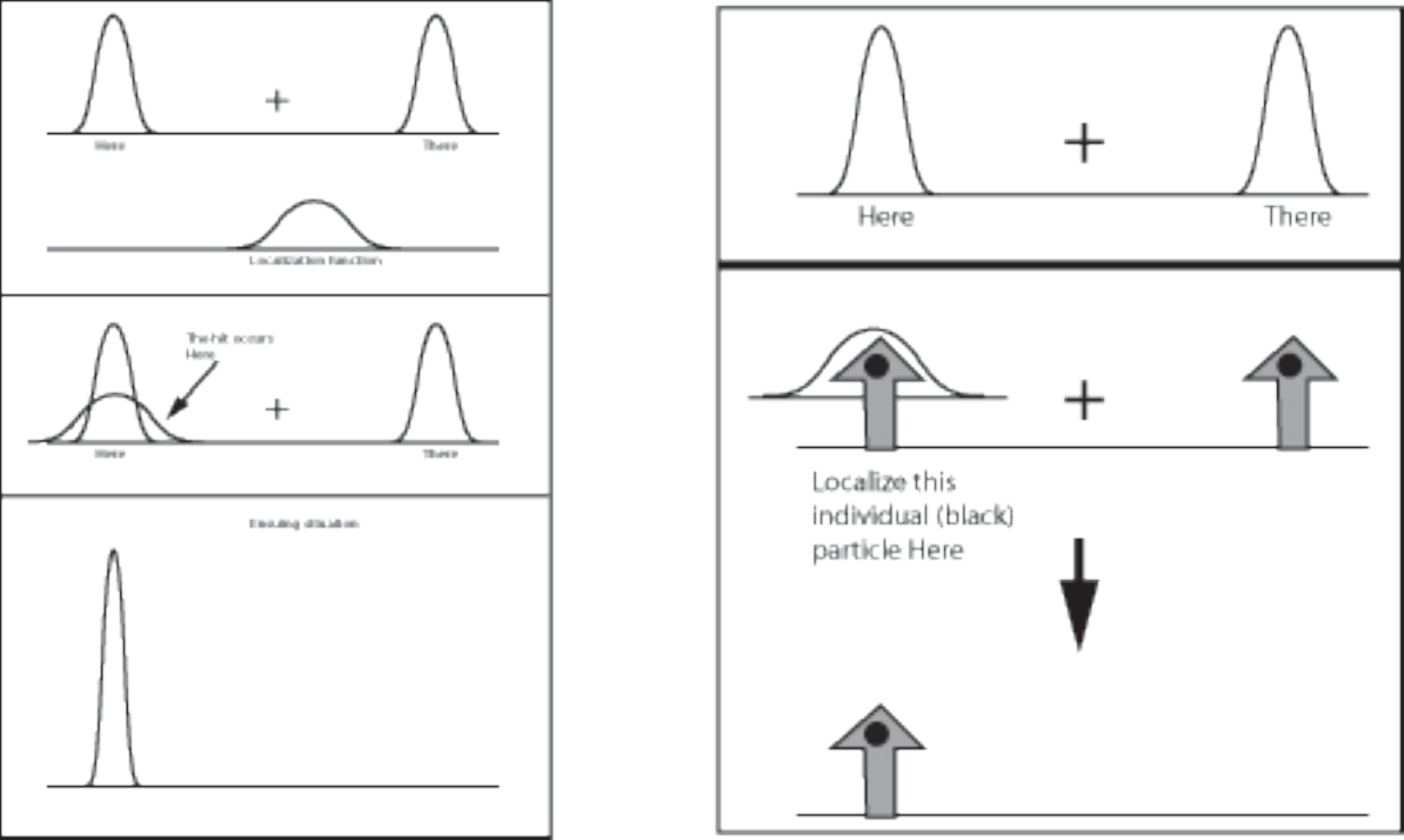}
\caption{The localization of a single particle and of a macro-object according to GRW.}
\end{figure}
With these premises we can now make the choice for the parameters $\alpha$ (note that $\frac{1}{\sqrt{\alpha}}$ gives the localization accuracy) and $\lambda$ of the theory. In ref.[5] we have chosen:
\begin{equation}
\alpha\simeq 10^{10}cm^{-2}, \lambda\simeq 10^{-16}sec^{-1}.
\end{equation}
Note that with these choices a microscopic system suffers a localization about every $10^{7}$ years, and this is why the theory agrees with quantum mechanics for such systems, a macroscopic body every $10^{-8} sec$ (due to the trigger mechanism in the case of an Avogadro number of particles the frequency becomes $10^{24}\cdot 10^{-16}=10^{8}$). As commented [6] by J. Bell: {\it The cat is not both dead and alive for more than a split second.}
  
The conclusion should be obvious. A universal dynamics has been worked out which leaves (practically) unaltered all quantum predictions for microsystems  but it accounts for wave packet reduction with probabilities in agreement with the quantum ones and for the classical behaviour of macroscopic systems, as well as for our definite perceptions concerning them.

Few remarks: i). The model has been generalized and formulated in a mathematically much more satisfactory but physically equivalent  way [46,47] by resorting to stochastic dynamical equations of the Ito or Stratonowich type, ii) The model is manifestly phenomenological but it gives some clear indications concerning the fact that the macro-objectification or measurement problem admits a consistent solution, iii). The model, even though it almost completely agrees  with quantum mechanics at the microscopic level qualifies itself as a rival theory of quantum mechanics, one which can be tested against it. Accordingly, it suggests where to look for putting into evidence possible violations of linearity. In recent years a lot of work in this direction is going on.

Obviously, also Collapse theories exhibit nonlocal features. However for them there is no theorem forbidding to get a generalization which does not resort to a preferred reference frame. Lot of work has been done along these lines; I will limit myself to mention some relevant steps. Before doing this, I consider it interesting to stress that the problem of having a theory inducing instantaneous reduction at-a-distance is a quite old one which has seen a lively debate and important contributions by Landau and Peierls [48], Bohr and Rosenfeld [49], Hellwig and Kraus [50] and Aharonov and Albert [51,52].

 Soon after the GRW theory has been formulated, P. Pearle [53] has presented a field theoretic relativistic generalization of it which has subsequently [54] been shown to be fully Lorentz invariant. Unfortunately, the model had some limitations arising from the occurrence of divergences which were not easily amendable. In 2000 the author of the present chapter has presented [55] a genuinely relativistic toy model of a theory inducing reductions. The model satisfies all the strict conditions identified in refs.[51,52]. F. Dawker and collaborators [56] have presented, in 2004, a relativistic collapse model on a discrete space that does not require a preferred slicing of space-time.
 
 The really important steps, however, occurred starting from 2007. R. Tumulka has presented [57] a fully satisfactory and genuinely relativistic invariant dynamical reduction model for a system of noninteracting fermions. Another important contribution [58] came from D. Bedingham. Finally, few month ago, a convincing proof of the viability of the Collapse theories in the relativistic domain has been presented [59]. The nice fact is that the conceptual attitude which underlies this attempt is that what the theory assumes to be true of the world around us is the mass density of the whole universe. In this way one recovers a unified, general picture of a quantum universe both at the micro and macro levels.
 
 I believe that the best way to conclude this Chapter, which has dealt in detail with the compatibility of quantum effects and relativistic requirements, is to quote a clarifying sentence from R. Tumulka [57], who has studied in great detail both the Bohmian as well as the Collapse approaches  to this fundamental problem:
 \begin{quote}
{\it A somewhat surprising feature of the present situation is that we seem to arrive at the following alternative: Bohmian mechanics shows that one can explain quantum mechanics, exactly and completely, if one is willing to pay with using a preferred reference slicing of space-time; our model suggests that one should be able to avoid a preferred slicing if one is willing to pay with a certain deviation from quantum mechanics.}
\end{quote}

{\bf Acknowledgements}

We thank Dr. R. Romano for an accurate critical reading of the manuscript.

\section{References}
[1]. Schr\"{o}dinger E. Discussion of Probability Relations between Separated Systems. Proceedings of the Cambridge Philosophical Society 1935; 31: 555-563.

[2]. Bell J.S. On the Einstein-Podolski-Rosen Paradox. Physics 1964; 1: 195-200 .

[3]. von Neumann J.  Matematische Grundlagen der Quantenmechanik. Berlin: Springer Verlag; 1932. (Engl. trans., Princeton University Press, 1955).

[4]. Herbert N. FLASH-- A Superluminal Communicator Based Upon a New Kind of Quantum Measurement. Foundations of Physics 1982; 12: 1171-79.

[5]. Ghirardi GC., Rimini A. and Weber T.   Unified Dynamics for Microscopic and Macroscopic Systems. Physical Review D 1986; 34: 470-491.

[6]. Bell J.S. Are there Quantum Jumps? in: Schr\"{o}dinger: Centenary Celebration of a Polymath. Cambridge: Cambridge University Press; 1987.

[7]. Bell J.S. Against Measurement in: Sixty-two Years of Uncertainty. New York: Plenum; 1990.

[8]. Bell JS. First and Second Class Difficulties in Quantum Mechanics. Journal of Physics A 2007; 40: 2921-2933.

[9]. Bassi A. and Ghirardi GC. A General Argument against the Universal Validity of the Superposition Principle. Physics Letters A 2000; 275: 373-381.

[10]. Bassi A. and Ghirardi GC. Dynamical Reduction Models. Physics Reports 2003;  379: 257-426.

[11]. Wigner EP. Die Messung Quantenmechanischer Operatoren. Zeitschrift f\"{u}r Physik 1952; 131: 101-108.

[12]. Araki H. and Yanase MM. Measurement of Quantum Mechanical Operators. Physical Review 1961; 120: 622-626. 

[13]. Yanase MM. Optimal Measuring Apparatus. Physical Review 1961; 123: 666-668.

[14]. Ghirardi GC.,  Miglietta F., Rimini A. and Weber T. Limitations on quantum measurements. I. Determination of the Minimal Amount of Nonideality and Identification of the Optimal Measuring Apparatuses. Physical Review D 1981; 24: 347-352.

[15]. Ghirardi GC., Rimini A. and Weber T. Quantum Evolution in the Presence of Additive Conservation Laws and the Quantum Theory of Measurement. Journal of Mathematical Physics 1982; 23: 1792-1796.

[16]. Kraus K. States, Effects and Operations, Berlin: Springer; 1983.

[17]. Kaiser D. How the Hippies Saved Physics, New York: W.W. Norton \& Co.; 2011.

[18]. Cufaro Petroni N.,  Garuccio A., Selleri F. and Vigier  JP. On a Contradiction between the Classical (idealized) Quantum Theory of Measurement and the Conservation of the Square of the Total Angular Momentum. C.R. Acad. Sci. Ser. B (Sciences Physiques)1980; 290: 111-114.

[19]. Selleri F. in: International Seminar on Mathematical Theory of Dynamical Systems and Microphysics. Udine; 1979. 

[20]. Herbert N. QUICK-a New Superluminal Transmission Concept. Boulder Creek, Cal.: C-Life Institute; 1979.

[21]. Beth R. Mechanical Detection and Measurement of the Angular Momentum of Light. Physical Review 1936; 50: 115-25.

[22]. Ghirardi GC. and Weber T. On Some Recent Suggestions of Superluminal Communication through the Collapse of the Wave Function. Lettere al Nuovo Cimento 1979;  26: 599-603.

[23]. Popper K. Quantum Theory and the Schism in Physics, London: Hutchinson; 1982.

[24]. Ghirardi GC., Marinatto L. and de Stefano F. Critical Analysis of Popper's Experiment. Physical Review A 2007; 75: 042107-1-5.

[25]. Ghirardi GC. Sneaking a Look at God's Cards, Princeton: Princeton University Press; 2005.

[26]. Ghirardi GC., Rimini A. and Weber T.  A General Argument against Superluminal Transmission through the Quantum Mechanical Measurement Process. Lett. Nuovo Cimento 1980; 27: 293-298.

[27]. Ghirardi GC., Grassi R., Rimini A. and Weber T. Experiments of the EPR Type Involving CP-Violation do not Allow Faster-than-light Communication between Distant Observers. Europhysics Letters 1988; 6: 95-100.

[28]. Eberhard PH. Bell's Theorem and the Different Concepts of Locality. Nuovo Cimento B 1978; 46: 392-419.

[29]. Shimony A. Controllable and Uncontrollable Nonlocality. in: Foundations of Quantum Mechanics  in the Light of New Technology. Tokyo: The physical Society of Japan;  1984.

[30]. Peres A. How the No-Cloning Theorem got its Name. Fortschritte der Physik 2003; 51: 458-61.

[31]. Ghirardi GC. and Weber T. Quantum Mechanics and Faster-than-Light Communication: Methodological Considerations. Il Nuovo Cimento B 1983; 78: 9-20.  

[32]. Wotters WK. and Zurek  WH. A Single Quantum Cannot be Cloned. Nature 1982; 299: 802-803.

[33]. Dieks D. Communication by EPR Devices. Physics Letters A 1982; 92: 271-272.

[34]. Wigner EP. The Probability of the Existence of a Self-Reproducing Unit, in: The Logic of Personal Knowledge. London: Routledge \& Kegan Paul, 1961.

[35]. Alicki R. Physical Limits on Self-Replication Processes. Open Systems and Information Dynamics 2006; 13: 113-117.

[36]. Eigen M. Selforganization of Matter and Evolution of Biological Macromolecules. Naturwissenschaften 1971; 58: 465-523.

[37]. Greenberger DM. If one Could Build a Macroscopic Schr\"{o}dinger Cat State one could Communicate Superluminally, in: Modern Studies of Basic Quantum Concepts and Phenomena. Singapore: World Scientific Publishing Co. 1998.

[38]. Kalamidas DA. A Proposal for a feasible quantum-optical experiment to test the validity of the no-signaling theorem, ArXiv:1110, 4269.

[39] Ghirardi GC. and Romano R. On a Proposal of Superluminal Communication. Journal of Physics A 2012; 45: 232001.

[40]. Shiekh AY. The Role of Quantum Interference in Quantum Computing. International Journal of Theoretical Physics 2006; 45: 1653-1655.

[41]. Bassi A. and Ghirardi. GC. On a Recent Proposal of Faster-than-Light Quantum Communication. International Journal of Theoretical Physics 2008 47: 2500-2506.

[42]. Bohm D. A Suggested Interpretation of the Quantum Theory in Terms of Hidden Variables. Physical Review 1952; 85: 166-193.

[43]. Ghirardi GC. and Grassi R. Bohm's Theory versus Dynamical Reduction, in: Bohmian Mechanics and Quantum Theory, an Appraisal. The Netherlands: Kluwer Academic Publishers. 1966. 

[44]. D\"{u}rr, D., Goldstein, S., M\"{u}nch-Berndl, K., Zanghi, N. Hypersurface Bohm-Dirac Models. Physical Review A 1999;  60: 2729-2736.

[45]. Bohm, D., Hiley, B. J. The Undivided Universe. London: Routledge (1993).

[46]. Pearle P. Combining Stochastic Dynamical State-Vector Reduction with Spontaneous Localization. Physical Review A 1999; 39: 2277-2289

[47]. Ghirardi GC., Pearle P. and Rimini A. Markov Processes in Hilbert Space and Continuous Spontaneous Localization of Systems of Identical Particles, Physical Review A 1990; 42: 78-89. 

[48]. Landau and Peierls R.  Erweiterung des Unbestimmtheitsprinzips f\"{u}r die relativistische Quantentheorie. Zeischrift f\"{u}r Physik, 1931; 69: 56-69.

[49]. Bohr N. and Rosenfeld L. Zur Frage der Messbarkeit der Electromagnetischen Feldgr\"{o}ssen, Kopenaghen. 1933.

[50]. Hellwig KE.  and Kraus K. Formal description of measurements in local quantum field theory. Physical Review D 1970; 1: 566-571.

[51]. Aharonov Y. and Albert DZ. States and observables in relativistic quantum field theories. Physical Review D 1980; 21: 3316-3324 ,  

[52]. Aharonov Y. and Albert DZ. Can we Make Sense out of the Measurement Process in Relativistic Quantum Mechanics? Physical Review D 1981; 24:359-370

[53]. Pearle P. Toward a Relativistic Theory of Statevector Reduction, in: Sixty-Two Years of Uncertainty. Plenum Press, New York; 1990.

[54]. Ghirardi GC., Grassi R. and Pearle P. Relativistic Dynamical Reduction Models: General Framework and Examples. Foundations of Physics 1990; 20: 1271-1316.

[55]. Ghirardi GC. Local Measurement of Nonlocal Observables and the Relativistic Reduction Process. Foundations of Physics  2000; 38: 1337-1385.

[56]. Dowker, F., Henson, J. Spontaneous Collapse Models on a Lattice. Journal of Statistical Physics 2004; 115: 1327-1339.

[57]. Tumulka, R. A Relativistic Version of the Ghirardi-Rimini- Weber Model.  Journal of Statistical Physics 2006; 125: 821-840.

[58]. Bedingham, DJ. Relativistic state reduction dynamics. ArXiv  1003-2774v2, 2010.

[59]. Bedingham  DJ., D\"{u}rr D., Ghirardi GC., Goldstein S., Tumulka R. and Zanghi. N. Matter Density and Relativistic Models of Wave Function Collapse. ArXiv 1111-1425, 2012.

\begin{figure}[htb]
\centering
\includegraphics[width=14cm]{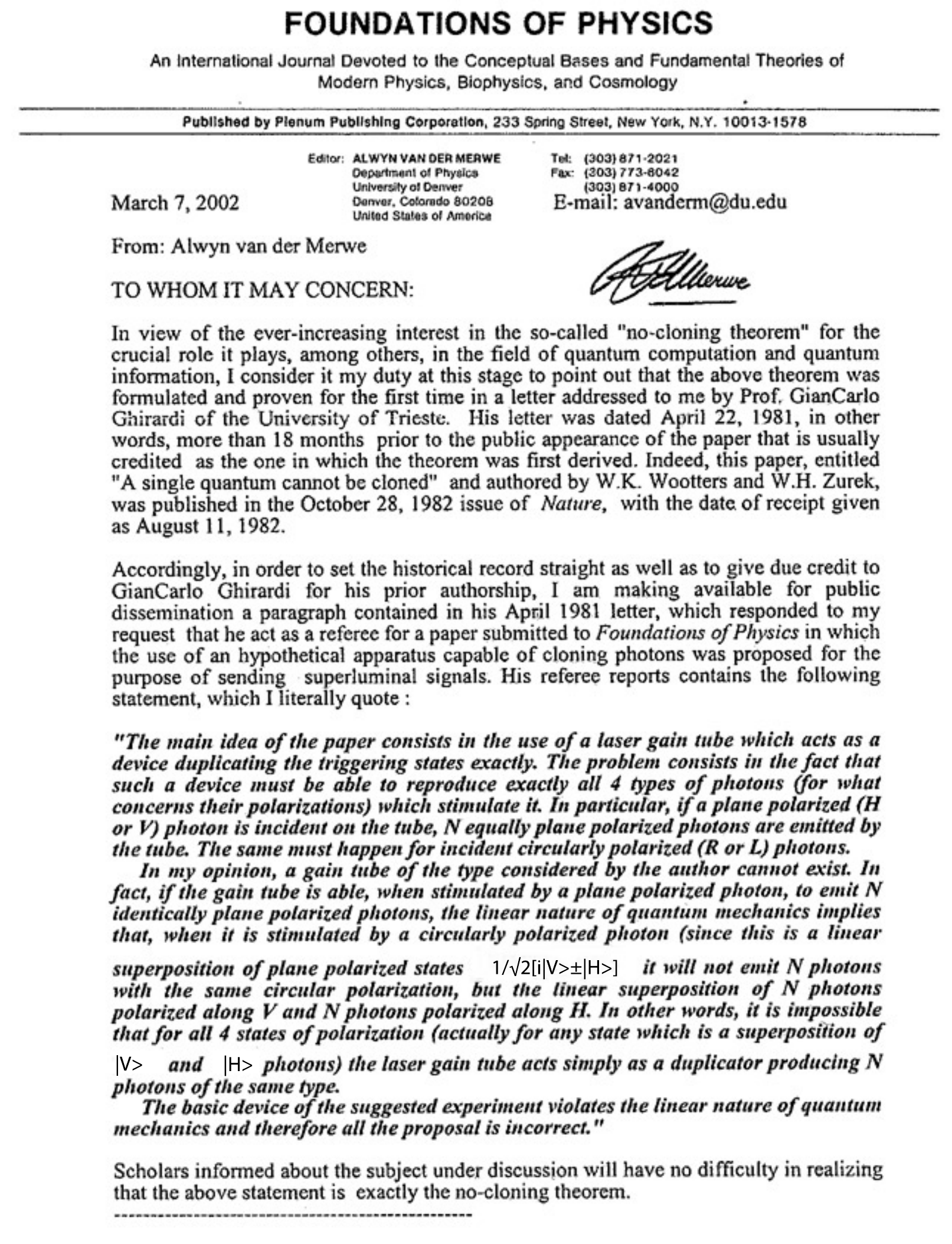}
\end{figure}

\end{document}